# Fail Fast - Fail Often: Enhancing Agile Methodology using Dynamic Regression, Code Bisector and Code Quality in Continuous Integration (CI)

**Sandeep Sivanandan***
Sr. Software Engineer, Security Technology Group, Cisco Systems India Pvt. Ltd., Bangalore, Karnataka, India

## Abstract

***Agile practices are receiving considerable attention from industry as an alternative to traditional software development approaches. However, there are a number of challenges in combining Agile [2] with Test-driven development (TDD) [10] practices, cloud deployments, continuous integration (CI), non-stop performance, load, security and accessibly testing. From these challenges; Continuous Integration is a relatively an approach widely discussed and practiced in software testing. This paper describes an approach for improved Agile Methodology using Code Quality, Code Bisector and Dynamic Regression in Continuous Integration. The set of tools used for this analysis, design and development are Jenkins, Robot Framework [4], Perforce and Git.***

## Keywords

***Testing; Automation Framework; Continuous Integration; Code Bisector; Code Quality; Regression; Robot Framework; Jenkins; Sonar; Corbetura.***

## 1. Introduction

Traditional software development methods didn't had process to solve how frequently or regularly will the code get integrated to entire source on a project. Software Developers used to work separately for hours, days, or even weeks on the same source without realizing how many conflicts (and perhaps defects) they are developing. In Agile; teams are producing robust code during each iteration, typically find that they are slowed down by the long diff-resolution and debugging sessions that often occur at the end of long integration cycles. The more developers are sharing the same code, the more problematic this is. For one of these reasons, agile teams often therefore choose to use Continuous Integration.

With a tool like Cruise Control or Jenkins and various source-control systems. Agile teams typically configure CI to include automated compilation, unit test execution, and code coverage and source control integration. Sometimes CI also includes automatically running automated acceptance tests In effect; CI means that the build is nearly always clean.

## 2. Problem Definition

There are many problems with current approaches, the paper describes below three problems:

a. Dynamic Regression: How can we run unit and functional automation for only the code changes happened?
b. Code Bisector: How faster can we intimate code breakages to the software developers?
c. Code Quality: How to define the quality of the code?

## 3. Dynamic Regression

As described in paper [8], in particular, at attempts to allocate trend component as a polynom the researcher collides with such infringements of circuits Regression Analysis (RA) [7] as: a high degree of correlation dependence between the subsequent and previous members of time lines as the rests; infringement of the assumption about normality of distributions of the rests, frequently caused by presence of regular displacement and a changeable dispersion.

Continuous Integration today as such is time consuming and sometimes in efficient for agile teams to move forward to Continuous Delivery. Some of the factors are:

*Author for correspondence

1. Jenkins [6] today monitors the source-control for code changes.
2. Starts a new build when there are changes committed by developer.
3. Triggers a full suite of thousands of Unit tests which takes hours of execution and might have regression failures.
4. Triggers an automated functional test job with ALL or some pre-defined sanity test cases which might not be related to the changes.

This percolates to inefficient testing and time to market. The Solution to the above problems was to build an Intelligent and Dynamic regression automation engine [9][7]. More about CI-Jenkins flow is explained along with the Flow diagram in figure 1.

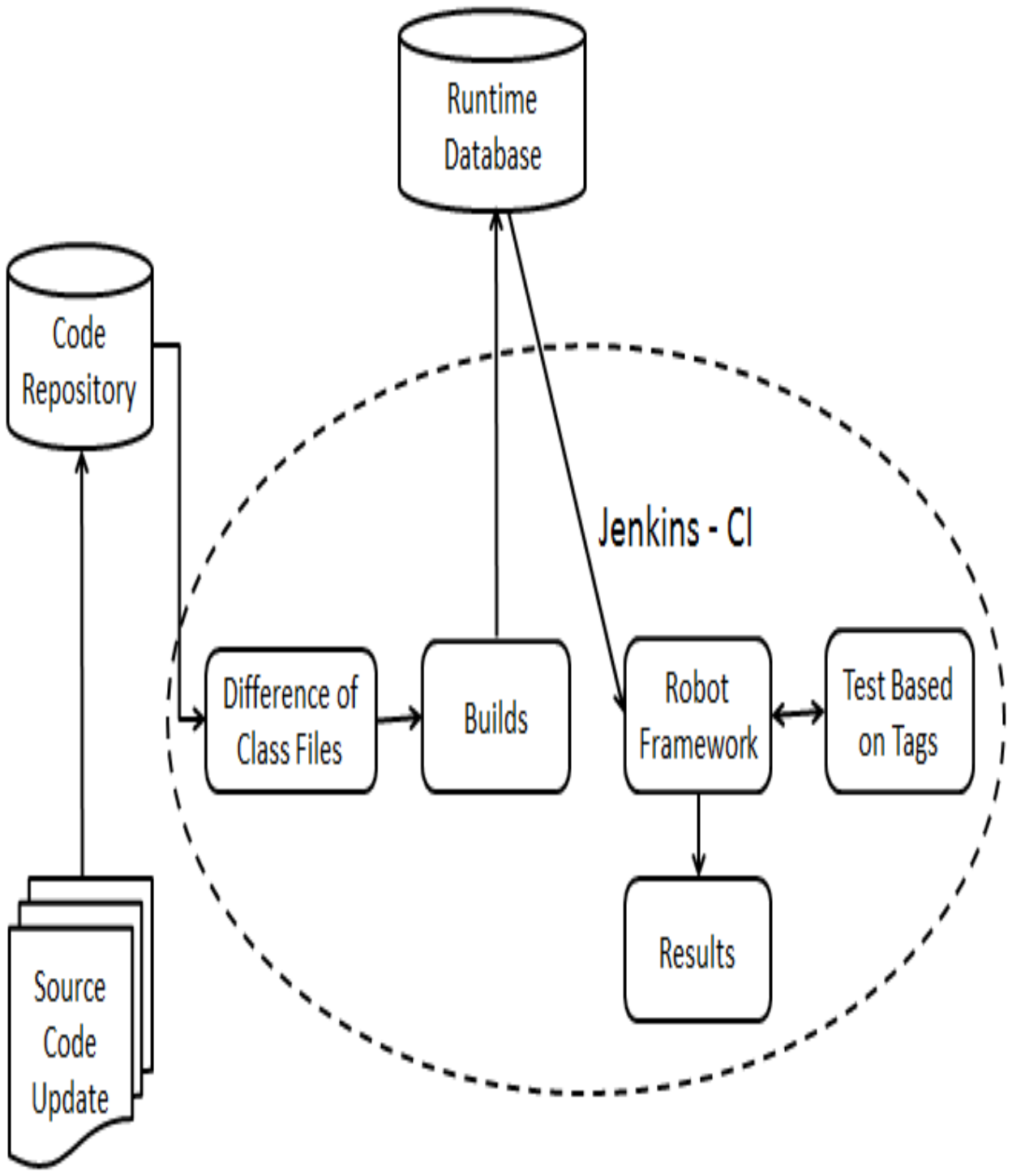

**Figure 1: Continuous Integration – Jenkins Based Dynamic Automation Engine**

Automated Dynamic Regression relies have multiple steps to make the system work.

First one was to have a mapping database, where source codes are mapped against the test suites. This is a cumbersome activity as thousands and lakhs of source code files needs to be mapped against the test suites.

**Step 1:**
Running code coverage for particular test suites and identifying the related classes which are called during the execution. The mapping would look something like below image:

**Table 1: Sample Table Mapper**

| Test Suite | Source Code Class |
|---|---|
| | Class A |
| | Class C |
| Class A | Class XF |
| | Class DE |
| | Class IM |

This is a humungous effort to build a mapper database the first time and intelligently adding any new Source code class and Test Suite to the Mapper database.

Smart Engine runs every 6 hours to fetch the list of files newly added to the source control system. It's been pushed to the database to compare when code coverage runs. If any new files are touched during this, the particular source code file is added to the mapper, incase removed or modified, it's taken care the same way. The mapper code takes care of not inserting configuration files, information files etc. to the mapper database. Once the code commit happens, the below steps are performed.

**Functional Automation:**
1. For the list of Source code files changed, the respective parser runs through the mapper database and pulls the list of corresponding test suites.
2. Once the list is generated, run automation for the particular test suites.

**Unit Test Automation:**
1. For the list of source code files changes, the parser runs through mapper database and pulls the list of corresponding test cases or test suites.
2. Once the list is generated, run automation for particular test suites or test cases.
3. Unit test automation with the above approach is much faster for knowing the

breakages on the code before its gets merged to the main branch for build.

## 4. Code Bisector

As part of Continuous Integration [6][11], a breakage of test becomes tougher to root cause. Apparently the delta changes can be more than a dozen files and to trace down at the end of the execution takes manual effort.

Software engineering practice of Code Bisector helps us to resolve this. "Bisection is a method used in software development to identify change sets that result in a specific behavior change. It is mostly employed for finding the patch that introduced a bug. Another application area is finding the patch that indirectly fixed a bug."

The same practice is followed and integrated to CI to intimate delta change breakages[6] to the development and fix then sooner and keep the build green. This information is blotted up in the quality dashboards with who are responsible for the breakages. Figure 2 shows the workflow of the code bisection mechanism in CI.

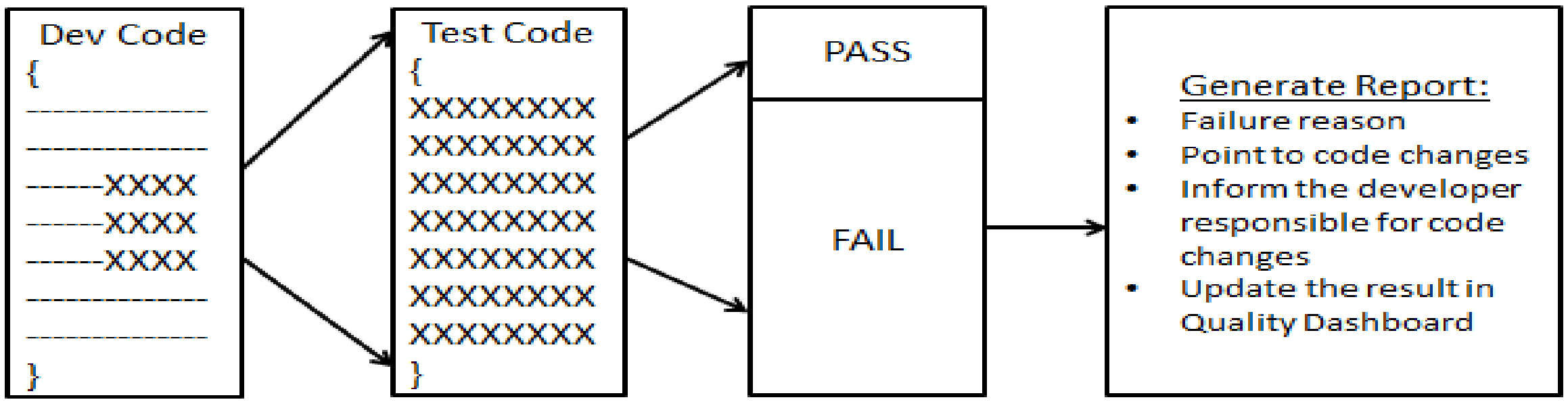

**Figure 2: CI – Code Bisection Mechanism**

## 5. Code Quality

Being agile [3], teams are drastically making changes to the code, and CI keep running unit test and automation to make sure the builds are green. But along with this how do we make sure the quality of the code written is good? How do we make sure quality of code doesn't break the basic security and performance benchmarks?

It's important to measure the quality of the code. Tools like Sonar; helps agile development teams to manage code quality efficiently.
As part of the code quality the seven axes which quality relies on:

- Coding standards and best practices.
- Code comments in the source code, especially in public APIs.
- Duplicate lines of code.
- Code Complexity amongst components.
- Zero or low code coverage by unit tests, especially in complex part of the program.
- Un Attending potential bugs
- Complex Design / Architecture.

**Sonar dashboard**
Figure 3: shows a sample of a Sonar Dashboard with above seven axes for code quality.

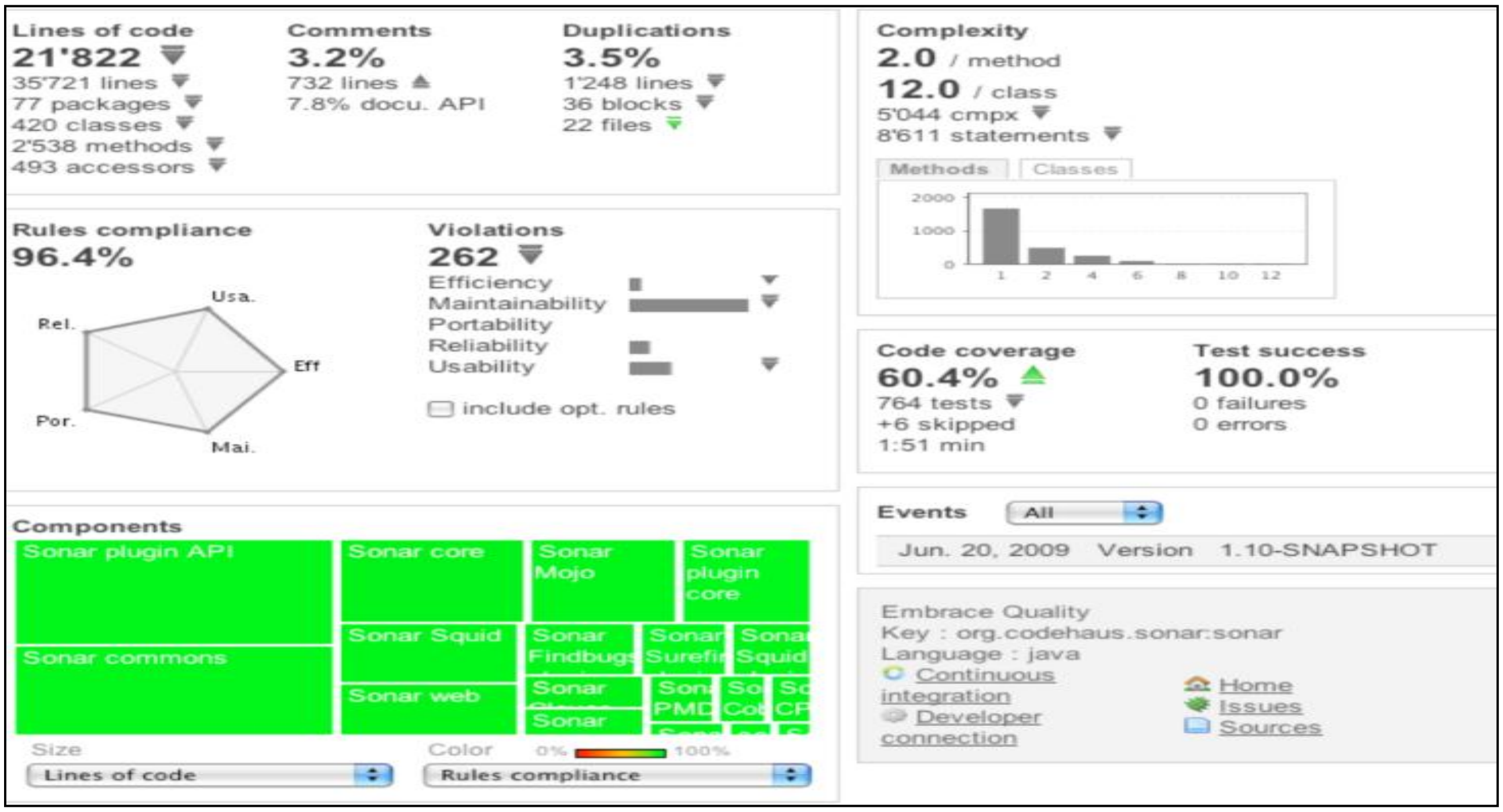

**Figure 3: Sonar Dashboard with seven axes for Code Quality**

As we talk more about the code quality, security is also very essential part. How are we taking care of Security?

Sonar have Security Rules Plugin, to identify the set of vulnerabilities like SQL Injection, Password Management, Error Handling and Logging mechanisms, Direct Object References etc. [5]. Tools like Findbugs helps find more security related during the time code is developed. A set of defined rules helps the developers to better code quality.

## 6. Benefits of using CI with Dynamic Regression, Code Bisector and Code Quality

All the above 3 process have helped the Business group to increase the Return on investment [ROI] by 70-80%, reduced bug finding time, code quality and On time delivery.

Dynamic regression evolved to a new generation where human intervention of validating delta changes to the source code is not needed. Automation automatically builds the suite and runs and exports results. Testers and automation teams are more concentrated on building business value and new areas.

Code Bisector has helped to fix the automation breakages and finding bugs faster. Regression changes are emailed and informed to the manger and respective developers.

Code quality is always updated Live, on quality dashboards helping the whole team aware. Helps the team to fix and solve the “Red” lines sooner. Builds quality and time to release to customer is much higher.

With above-mentioned techniques, predictive quality can be attained.

## 7. Conclusion and Future work

As Agile practice are receiving considerable attention from industry as an alternative to traditional software development techniques. There are multiple challenges in testing domain. The paper exposed an enhancement to existing Agile methodology to improve the challenges and difficulties faces in the development life cycle. The new approaches:

Dynamic regression, Code Bisector and Code Quality techniques are discussed and from Dynamic Regression observed that Unit test automation with the above approach is much faster for knowing the breakages on the code before its gets merged to the main branch for build. Code bisector technique is focused on the practice that followed and integrated to CI to intimate delta change breakages to the development and fix then sooner and keep the build green. Code Quality explained the how to retain the quality of the code. Tools like Sonar; used to showcase the results obtained in agile development teams to manage code quality efficiently.

The results obtained from these techniques will be compare with the existing techniques like Tag based testing, Generic Water Fall Techniques, Spiral Approach, Round Robin Approach etc. Will be improvised and adapted for different types of testing including hardware and Non- GUI [Commands line Interface applications].

## Acknowledgment

The author wish to thank members of Test Automation–ESA; Security Technology Group (STG), Cisco Systems, Bangalore for their help in completing this work.

## References

[1] John Walker, "Beyond Scrum: Continuous Integration with Build and Test Automation", Online: www.perforce.com/company/newsletter/2013/04/beyond-scrum-continuous-integration-build-and-test-automation, Perforce Software Company News Letter, S No. 04, 2013.
[2] Bhanu Prakash G, Yogeesha C B and Prabhu Periasamy, "Highly scalable model for tests execution in cloud environments", 20th annual International Conference on Advanced Computing and Communications (ADCOM 2012) at Bangalore, 14th – 16th Dec 2012.
[3] Crispin, Lisa, and Janet Gregory. Agile testing: A practical guide for testers and agile teams. Pearson Education, 2009.
[4] Nokia Solutions and Networks, "Robot Framework User Guide, Version 2.8.5", online: http://robotframework.googlecode.com/hg/doc/userguide/RobotFrameworkUserGuide.html?r=2.8.5, Section 1 - 4, 2014.
[5] Sarah Goff-Dupont, "A skeptic's guide to continuous delivery, part 2: the nuts & bolts of CI", Online: http://blogs.atlassian.com/2014/07/skeptics-guide-continuous-delivery-part-2-nuts-bolts-ci/, Atlassian Blogs Product update, S No. 7, 2014.
[6] Kohsuke, "Jenkins: An extendable open source continuous integration server", Online Tutorial: http://jenkins-ci.org/category/tags/tutorial, Jenkins Tutorial Material, 2014.
[7] S.G.Valeev, "Techique, Algorithms and the Software of Dynamic Regression Modelling", Microsymposium 38, MS091, 2003, http://www.planetary.brown.edu/planetary/international/Micro_38_Abs/ms091.pdf .
[8] Sandeep Sivanandan and Yogeesha C B, "Agile Development Cycle: Approach to Design an Effective Model Based Testing with Behaviour Driven Automation Framework", 20th annual International Conference on Advanced Computing and Communications (ADCOM 2014) at Bangalore, 19th - 22nd September 2014.
[9] Yogeesha, C. B., Ramachandra V. Pujeri, and R. S. Veena. "Randomized Algorithms: On the Improvement of Searching Techniques Using Probabilistic Linear Linked Skip Lists." Proceedings of International Conference on Advances in Computing. Springer India, 2012.
[10] Lehman, Tobin J., and Akhilesh Sharma. "Software development as a service: agile experiences." SRII Global Conference (SRII), 2011 Annual. IEEE, 2011.
[11] Christian Couder, "Fighting regressions with git bisect", Online: https://www.kernel.org/pub/software/scm/git/docs/git-bisect-lk2009.html, The Linux Kernel Archives, GIT Repository, Version 4.0.5, Folder 3, 2009.

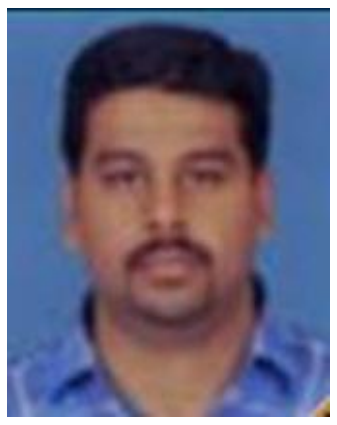

**Sandeep Sivanandan** born in Kerala, India, in 1983. He is an M.Tech from SAM Higginbottom Institute, Allahabad. Over 11 years of rich experience in Testing, Test Automation, Automation Architecting, Agile, Mobile Technology, Virtualization, Security and Vulnerability assessments. Currently he is working as Sr.Software Engineer, with roles of Test Automation Lead in Cisco Systems India Pvt. Ltd. for Email Security, Bangalore from 2011.
Email: sasivana@cisco.com